\pdfoutput=1 
\documentclass[english]{lni}

\IfFileExists{latin1.sty}{\usepackage{latin1}}{\usepackage{isolatin1}}
\usepackage{graphicx}
\usepackage{amsfonts,amssymb}
\usepackage[fleqn]{amsmath}
\usepackage{subfigure}
\usepackage{alltt}
\usepackage{caption}
\usepackage[vlined,ruled,linesnumbered,commentsnumbered]{algorithm2e}
\usepackage{appendix}

\DeclareRobustCommand{\myTitle}{Parallel Sorted Neighborhood Blocking with MapReduce}

\title{\myTitle}

\author{Lars Kolb, Andreas Thor, Erhard Rahm\\\\
	Department of Computer Science, University of Leipzig, Germany \\
	$\lbrace$kolb, thor, rahm$\rbrace$@informatik.uni-leipzig.de}

\begin{document}

\maketitle

\begin{abstract}
Cloud infrastructures enable the efficient parallel execution of data-intensive tasks such as entity resolution on large datasets. We investigate challenges and possible solutions of using the MapReduce programming model for parallel entity resolution. In particular, we propose and evaluate two MapReduce-based implementations for Sorted Neighborhood blocking that either use multiple MapReduce jobs or apply a tailored data replication.
\end{abstract}

\section{Introduction}

Cloud computing has become a popular paradigm for efficiently processing data and computationally intensive application tasks \cite{cloud_computing}. Many cloud-based implementations utilize the MapReduce programming model for parallel processing on cloud infrastructures with up to thousands of nodes \cite{map_reduce}. The broad availability of MapReduce distributions such as Hadoop makes it attractive to investigate its use for the efficient parallelization of data-intensive tasks. 

Entity resolution (also known as object matching, deduplication, or record linkage) is such a data-intensive and performance-critical task that can likely benefit from cloud computing. Given one or more data sources, entity resolution is applied to determine all entities referring to the same real world object \cite{sorted_neighborhood, entity_resolution_2}. It is of critical importance for data quality and data integration, e.g., to find duplicate customers in enterprise databases or to match product offers for price comparison portals.

Many approaches and frameworks for entity resolution have been proposed \cite{er_book, er_survey, er_frameworks, er_learning}. The standard (naive) approach to find matches in $n$ input entities is to apply matching techniques on the Cartesian product of input entities. However, the resulting quadratic complexity of $O(n^2)$ results in intolerable execution times for large datasets \cite{er_frameworks2}. So-called blocking techniques \cite{fast_blocking} thus become necessary to reduce the number of entity comparisons whilst maintaining match quality. This is achieved by semantically partitioning the input data into blocks of similar records and restricting entity resolution to entities of the same block. Sorted neighborhood (SN) is one of the most popular blocking approaches \cite{sorted_neighborhood}. It sorts all entities using an appropriate blocking key and only compares entities within a predefined distance window $w$. The SN approach thus reduces the complexity to $O(n\cdot w)$ for the actual matching. 

In this study we investigate the use of MapReduce for the parallel execution of SN blocking and entity resolution. By combining the use of blocking and parallel processing we aim at a highly efficient entity resolution implementation for very large datasets. The proposed approaches consider specific partitioning requirements of the MapReduce model and implement a correct sliding window evaluation of entities. Our contributions can be summarized as follows:

\begin{itemize}
  \item We demonstrate how the MapReduce model can be applied for the parallel execution of a general entity resolution workflow consisting of a blocking and matching strategy.
  \item We identify the major challenges and propose two approaches for realizing Sorted Neighborhood Blocking on MapReduce. The approaches (called JobSN and RepSN) either use multiple MapReduce jobs or apply a tailored data replication during data redistribution. 
  \item We evaluate both approaches and demonstrate their efficiency in comparison to the sequential approach. The evaluation also considers the influence of the window size and data skew. 
\end{itemize}

The rest of the paper is organized as follows. In the next section we introduce the MapReduce programming paradigm. Section~\ref{sec_er_mr} illustrates the general realization of entity resolution using MapReduce. In Section \ref{sec_sn}, we describe how the SN blocking strategy can be realized based on MapReduce. Section~\ref{sec_experiments} describes the performed experiments and evaluation. Related work is discussed in Section~\ref{sec_related_work} before we conclude.

\section{MapReduce}
\label{sec_mr}

MapReduce is a programming model introduced by Google in 2004 \cite{DBLP:conf/osdi/DeanG04}. It supports parallel data-intensive computing in cluster environments with up to thousands of nodes. A MapReduce program relies on data partitioning and redistribution. Entities are represented by $(key,value)$ pairs. A computation is expressed with two user defined functions:
 
\begin{align*}
map&:(key_{in},value_{in}) \to list(key_{tmp},value_{tmp})\\
reduce&:(key_{tmp},list(value_{tmp})) \to list(key_{out},value_{out})
\end{align*}

These functions contain sequential code and are executed in parallel across many nodes utilizing present data parallelism. MapReduce nodes run a fixed number of mapper and/or reducer processes. Mapper processes scan disjoint input partitions in parallel and transform each entity in a $(key,value)$-representation before the $map$ function is executed. The output of a $map$ function is sorted by key and repartitioned by applying a partitioning function on the key. A partition may contain different keys but all values with the same key are in the same partition. The partitions are redistributed, i.e., all $(key,value)$ pairs of a partition are sent to exactly one node. The receiving node hosts a fixed number of reducer processes whereas a single reducer is responsible for handling the $map$ output pairs from all mappers that share the same key. Since the number of keys within a dataset is in general much higher than the number of reducers, a reducer merges all incoming $(key,value)$ pairs in a sorted order by their intermediate keys. In the reduce phase the reducer passes all values with the same key to a $reduce$ call.

\begin{figure}[t]
  \begin{center}
    \includegraphics[scale=0.39]{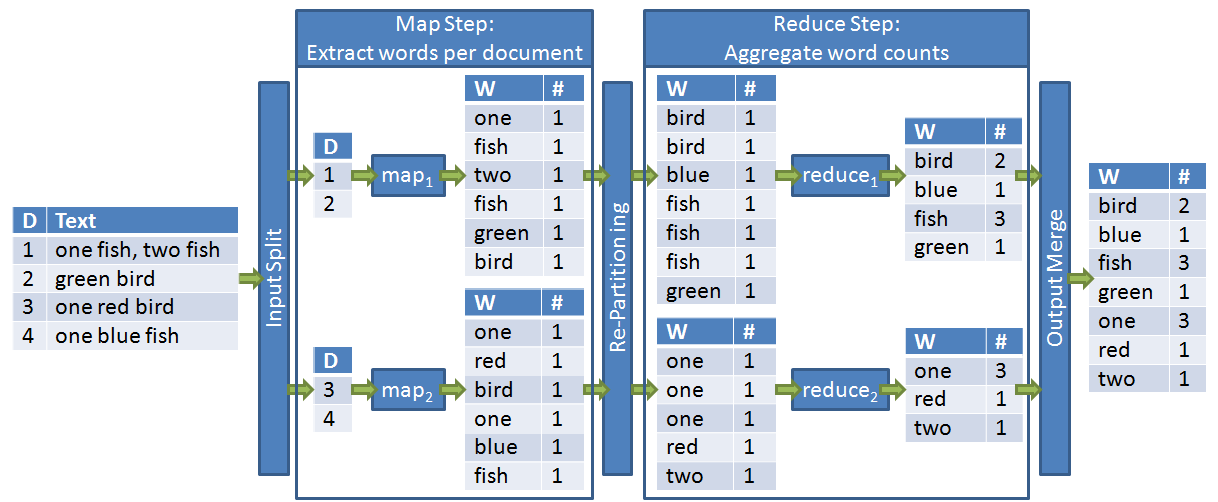}
    \caption{\label{fig_mapreduce}Example of a MapReduce program for counting word occurrences in documents (similar to \cite{jimmy_mr_book})}
  \end{center}
\end{figure}

An exemplary data flow of a MapReduce computation is shown in Figure~\ref{fig_mapreduce}. The MapReduce program counts the number of term occurrences across multiple documents which is a common task in information retrieval. The input data (list of documents) is partitioned and distributed to the $m=2$ mappers. In the simple example of Figure~\ref{fig_mapreduce}, two documents are assigned to each of the two mappers. However, a mapper usually process larger partitions in practice. Instances of the map function are applied to each partition of the input data in parallel. In our example, the map function extract all words for all documents and emits a list of $(term, 1)$ pairs. The partitioning assigns every ($key$,$value$) pair to one reducer according to the key. In the example of Figure~\ref{fig_mapreduce} a simple range partitioning is applied. All keys (words) starting with a letter from $a$ through $m$ are assigned to the first reducer; all other keys are transferred to the second reducer. The input partitions are sorted for all reducers. The user-defined reduce function then aggregates the word occurrences and outputs the number of occurrences per word. The output partitions of reduce can then easily be merged to a combined result since two partitions do not share any key.

There are several frameworks that implement the MapReduce programming model. Hadoop \cite{hadoop} is the most popular implementation throughout the scientific community. It is free, easy to setup, and well documented. We therefore implemented and evaluated our approaches with Hadoop. Most MapReduce implementations utilize a distributed file system (DFS) such as the Hadoop distributed file system \cite{hdfs}. The input data is initially stored partitioned, distributed, and replicated across the DFS. Partitions are redistributed across the DFS in the transition from map to reduce. The output of each reduce call is written to the DFS.

\section{Entity resolution with MapReduce}
\label{sec_er_mr}

In this work we consider the problem of entity resolution (deduplication) within one source. The input is a data source $S = \lbrace e_i \rbrace$ that contains a finite set of entities $e_i$.  The task is to identify all pairs of entities $M=\lbrace(e_i, e_k)\;|\; e_i, e_k \in S\rbrace$ that are regarded as duplicates.  

\begin{figure}[t]
  \begin{center}
    \includegraphics[scale=0.5]{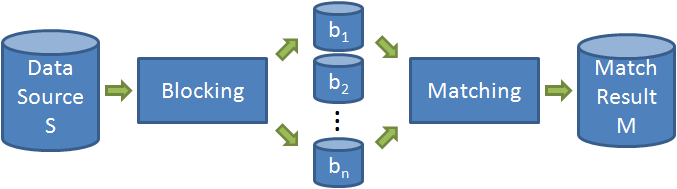}
    \caption{\label{fig_blocking}Simplified general entity resolution workflow}
  \end{center}
\end{figure}

Figure~\ref{fig_blocking} shows a simplified generic entity resolution workflow. The workflow consists of a blocking strategy and a matching strategy. Blocking semantically divides a data source $S$ into possibly overlapping partitions (blocks) $b_i$, with $S=\bigcup b_i$. The goal is to restrict entity comparison to pairs of entities that reside in the same block. The partitioning into blocks is usually done with the help of blocking keys based on the entities' attribute values. Blocking keys utilize the values of one or several attributes, e.g., product manufacturer (to group together all products sharing the same manufacturer) or the combination of manufacturer and product type. Often, the concatenated prefixes of a few attributes form the blocking key. A possible blocking key for publications could be the combination of the first letters of the authors' last names and the publication year (similar to the reference list in this paper). 

The matching strategy identifies pairs of matching entities of the same block. Matching is usually realized by pairwise similarity computation of entities to quantify the degree of similarity. A matching strategy may also employ several matchers and combine their similarity scores. As a last step the matching strategy classifies the entity pairs as match or non-match. Common techniques include the application of similarity thresholds, the incorporation of domain-specific selection rules, or the use of training-based models. Our entity resolution model abstracts from the actual matcher implementation and only requires that the matching strategy returns the list of matching entity pairs.  

The realization of the general entity resolution workflow with MapReduce is relatively straightforward by implementing blocking within the map function and by implementing matching within the reduce function. To this end, map first determines the blocking key for each entity. The MapReduce framework groups entities with the same blocking key to blocks and redistributes them. The reduce step then matches the entities within one block. Such a procedure shares similarities with the join computation in parallel database systems \cite{parallel_database}. There, the join key (instead of the blocking key) is used for data repartitioning to allow a subsequent parallel join (instead of match) computation. The join (merge) results are disjoint by definition and can thus easily merged to obtain the complete result.

\begin{figure}[t]
  \begin{center}
    \includegraphics[scale=0.45]{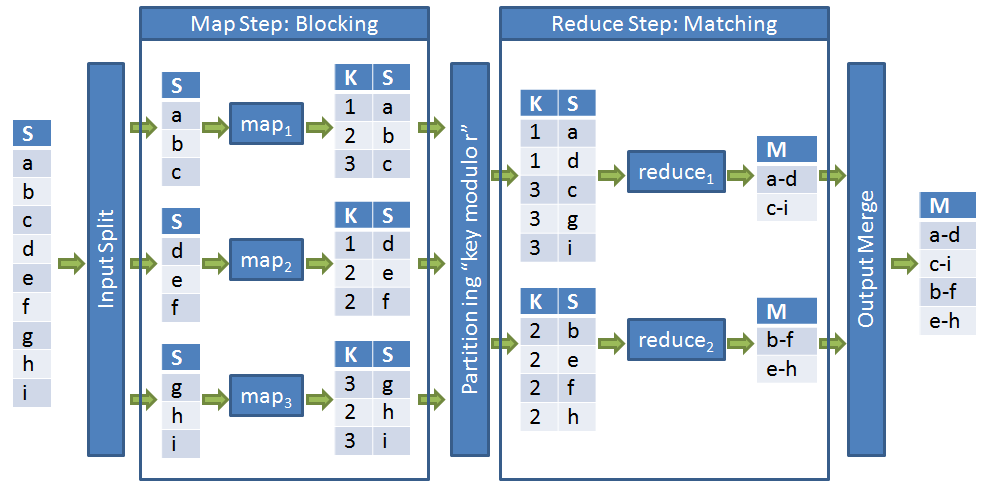}
    \caption{\label{fig_blocking_mr}Example of a general entity resolution workflow with MapReduce ($n=9$ input entities, $m=3$ mappers, $r=2$ reducers)}
  \end{center}
\end{figure}

Figure~\ref{fig_blocking_mr} illustrates an example for $|S|=n=9$ entities, $a$-$i$, of an input data source $S$ using $m=3$ mappers and $r=2$ reducers. First, the input partitioning (split) divides the input source $S$ into $m$ partitions and assigns one partition to each mapper. Then, the individual mappers read their (preferably) local data in parallel and determine a blocking key value $K$ for each of the input entities.\footnote{Figure~\ref{fig_blocking_mr} omits the map input keys for simplicity.} For example, entity $a$ has blocking key value $1$. Afterwards all entities are dynamically redistributed by a partition function such that all entities with the same blocking key value are sent to the same reducer (node). In the example of Figure~\ref{fig_blocking_mr}, blocking key values $1$ and $3$ are assigned to the first reducer whereas key $2$ is assigned to the second node. The receivers group the incoming entities locally and identify the duplicates in parallel. For example, the first reducer identifies the duplicate pairs $(a,d)$ and $(c,i)$. The reduce outputs can finally be merged to achieve the overall match result.

Unfortunately, the sketched MapReduce-based entity resolution workflow has several limitations:

\begin{description}
  \item[Disjoint data partitioning:] MapReduce uses a partition function that determines a single output partition for each map output pair based on its key value. This approach is suitable for many blocking techniques but complicates the realization of blocking approaches with overlapping blocks. For example, the sorted neighborhood approach does not necessarily only compare entities sharing the same blocking key.
  \item[Load balancing:] Blocking may lead to partitions of largely varying size due to skewed key values. Therefore the execution time may be dominated by a single or a few reducers. Load balancing and skew handling is a well-known problem in parallel database systems \cite{practial_skew_handling}. The adaptation of those techniques to the MapReduce paradigm is beyond the scope of this paper and left as a subject for future work. 
  \item[Memory bottlenecks:] All entities within the same block are passed to a single reduce call using an iterator. The reducer can only process the data row-by-row similar to a forward SQL cursor. It does not have any other options for data access. On the other hand, the matching requires that all entities within the same reduce block are compared with each other. The reducer must therefore store all entities in main memory (or must make use of other external memory) which can lead to serious memory bottlenecks. The memory bottleneck problem is partly related to the load balancing problem since skewed data may lead to large blocks which tighten the memory problem. Possible solutions have been proposed in \cite{setsim_mr}. However, memory issues can also occur with a (perfect) uniform key distribution.
\end{description}

In this work we focus on the first challenge and propose two approaches how the popular and efficient Sorted Neighborhood blocking method SN can be realized within a MapReduce framework. As we will discuss, the SN approach is less affected by load balancing problems. Moreover, the risk for memory bottlenecks is reduced since the row-by-row process matches the SN's sliding window approach very well.

\section{Sorted Neighborhood with MapReduce}
\label{sec_sn}

Sorted neighborhood (SN) \cite{sorted_neighborhood} is a popular blocking approach that works as follows. A blocking key $K$ is determined for each of $n$ entities. Typically the concatenated prefixes of a few attributes form the blocking key. Afterwards the entities are sorted by this blocking key. A window of a fixed size $w$ is then moved over the sorted records and in each step all entities within the window, i.e., entities within a distance of $w-1$, are compared.

\begin{figure}[t]
  \begin{center}
    \includegraphics[scale=0.45]{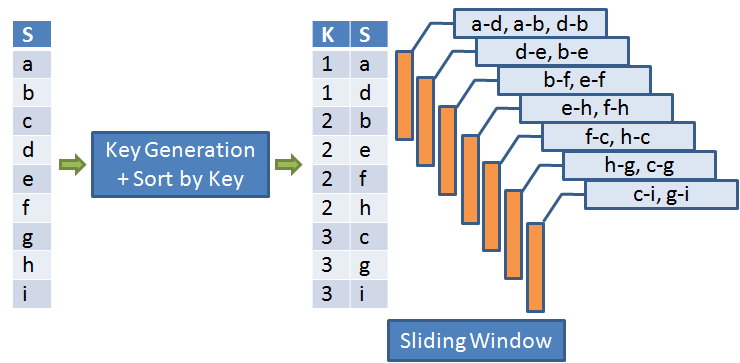}
    \caption{\label{fig_sorted_neighborhood}Example execution of sorted neighborhood with window size $w=3$}
  \end{center}
\end{figure}

Figure~\ref{fig_sorted_neighborhood} shows a SN example execution for a window size $w=3$. The input set consists of the same $n=9$ entities that have already been employed in the example of Figure~\ref{fig_blocking_mr}. The entities ($a$-$i$) are first sorted by their blocking keys (1, 2, or 3). The sliding window then starts with the first block $(a,d,b)$ resulting in the three pairs $(a,d)$, $(a,b)$, and $(d,b)$ for later comparison. The window is then moved by one step to cover the block $(d,b,e)$. This leads to two additional pairs $(d,e)$ and $(b,e)$. This procedure is repeated until the window has reached the final block $(c,g,i)$. Figure~\ref{fig_sorted_neighborhood} lists all pairs generated by the sliding window. In general, the overall number of entity comparisons is $(n-w/2)\cdot(w-1)$.

The SN approach is very popular for entity resolution due to several advantages. First, it reduces the complexity from $O(n^2)$ (matching $n$ input entities without blocking) to $O(n)+O(n\cdot \log n)$ for blocking key determination and sorting and $O(n\cdot w)$ for matching. Thereby matching large datasets becomes feasible and the window size $w$ allows for a dedicated control of the runtime. Second, the SN approach is relatively robust against a suboptimal choice of the blocking key since it is able to compare entities with a different (but similar) blocking key. The SN approach may also be repeatedly executed using different blocking keys. Such a multi-pass strategy diminishes the influence of poor blocking keys (e.g., due to dirty data) whilst still maintaining the linear complexity for the number of possible matches. Finally the linear complexity makes SN more robust against load balancing problems, e.g., if many entities share the same blocking key.

The major difference of SN in comparison to other blocking techniques is that a matcher does not necessarily only compare entities sharing the same blocking key. For example, entities $d$ and $b$ have different blocking keys but need to be compared according to the sorted neighborhood approach (see Figure~\ref{fig_sorted_neighborhood}). On the other hand, one of the key concepts of MapReduce is that map input partitions are processed independently. This allows for a flexible parallelization model but makes it challenging to group together entities within a distance of $w$ since a mapper has no access to the input partition of other mappers.

Even if we assume that a mapper can determine the relevant entity sets for each entity\footnote{For example, this could be realized by employing a single mapper only.}, the general approach as presented in Section~\ref{sec_er_mr} is not suitable. This is due to the fact that the sliding window approach of SN leads to heavily overlapping entity sets for later comparison. In the example of Figure~\ref{fig_sorted_neighborhood}, the sliding window produces the blocks  $\lbrace a, d, b \rbrace$ and $\lbrace d, b, e\rbrace$ among others. The general MapReduce-based entity resolution approach  is, of course, applicable, but would expend unnecessary resources. First of all, almost all entities appear in $w$ blocks and would therefore appear $w$ times in the map output. Finally, the overlapping blocks would cause the generation of duplicate pairs in the reduce step, e.g., $(d,b)$ in the above mentioned example.

We therefore target a more efficient MapReduce-based realization of SN and, thus, adapt the approach described in Section~\ref{sec_er_mr}. The map function determines the blocking key for each input entity independently. The map output is then distributed to multiple reducers that implement the sliding window approach for each reduce partition. For example, in the case of two reducers one may want to send all entities of Figure~\ref{fig_sorted_neighborhood} with blocking key $\leq 2$ to the first reducer and the remaining entities to the second reducer. The analysis of this scenario reveals that we have to solve mainly two challenges to implement a MapReduce-based SN approach.

\begin{description}

\item[Sorted reduce partitions:] 
The SN approach assumes an ordered list of all entities based on their blocking keys. A repartitioning must therefore preserve this order, i.e., the map output has to make sure that all entities assigned to reducer $R_x$ have a smaller (or equal) blocking key than all entities of reducer $R_{x+1}$. This allows each reducer to apply the sliding window approach on its partition. We will address the sorted data repartitioning by employing a composite key approach that relies on a partition prefix (see Section~\ref{sec_LocSN}). 

\item[Boundary entities:]
The continuous sliding window of SN requires that not only entities within a reduce partition but also across different reduce partitions have to be compared. More precisely, the highest $v<w$ entities of a reduce partition $R_x$ need to be compared with the $w-v$ smallest entities of the succeeding partition $R_{x+1}$. In the following, we call those entities \emph{boundary entities}. We propose two approaches (JobSN and RepSN) that employ multiple MapReduce computation steps and data replication, respectively, to process boundary entities and, thus, to map the entire SN algorithm to a MapReduce computation (Sections~\ref{sec_JobSN} and \ref{sec_RepSN}).  

\end{description}

\subsection{Sorted Reduce Partitions}
\label{sec_LocSN}

\begin{figure}[t]
  \begin{center}
    \includegraphics[scale=0.45]{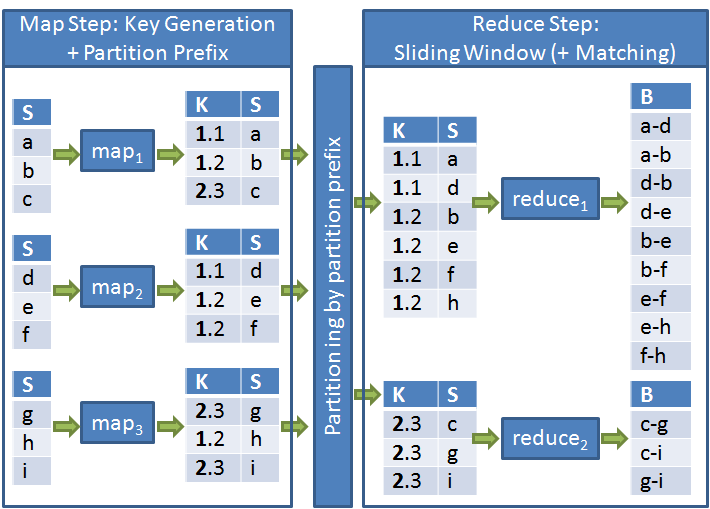}
    \caption{\label{fig_sn_mr}Example execution of sorted data partitioning with a composite key consisting of a blocking key and a partition prefix. The composite key ensures that the reduce partitions are ordered. If the sliding window approach ($w=3$) is applied to both reduce partitions, it is only able to identify 12 out of the 15 SN correspondences (as shown in Figure~\ref{fig_sorted_neighborhood}). The pairs $(f,c)$, $(h,c)$, and $(h,g)$ can not be found since the involved entities reside in different reduce partitions.}
  \end{center}
\end{figure}

We achieve sorted reduce partitions (SRP) by utilizing an appropriate user-defined function $p$ for data redistribution among reducers in the map phase. Data redistribution is based on the generated blocking key $k$, i.e., $p$ is a function $p: k \rightarrow i$ with $1 \leq i \leq r$ and $r$ is the number of reducers. A monotonically increasing function $p$ (i.e., $p(k_1) \geq p(k_2)$ if $k_1 \geq k_2$) ensures that all entities assigned to reducer $i$ have a smaller or equal blocking key than any entity processed by reducer $i+1$.

The range of possible blocking key values is usually known beforehand for a given dataset because blocking keys are typically derived from numeric or textual attribute values. In practice simple range partitioning functions $p$ may therefore be employed.

The execution of SRP is illustrated in Figure~\ref{fig_sn_mr} for $m=3$ mappers and $r=2$ reducers. It uses the same entities and blocking keys as the example of Figure~\ref{fig_sorted_neighborhood}. In this example the function $p$ is defined as follows: $p(k)=1$ if $k\leq 2$, otherwise $p(k)=2$. The map function first generates the blocking key $k$ for each input entity and adds $p(k)$ as a prefix. In the example of Figure~\ref{fig_sn_mr}, the blocking key value for $c$ is 3 and $p(k)=2$. This results in a combined key value 2.3. The partitioning then distributes the ($key$,$value$) pairs according to the partition prefix of the key. For example, all keys starting with 2 are assigned to the second reducer. Moreover, the input partitions for each reducer are sorted by the (combined) key. Since all keys of reducer $i$ start with the same prefix $i$, the sorting of the keys is practically done based on the actual blocking key.

Afterwards the reducer can run the sliding window algorithm and, thus, generates the correspondences of interest. Figure~\ref{fig_sn_mr} illustrates the resulting correspondences as reduce output ($B$=Blocking). For entity resolution the reduce function will apply a matching approach to the correspondences. Reduce will therefore likely return a small subset of $B$. However, since we investigate in blocking techniques we leave $B$ as output to allow for comparison with other approaches (see Section~\ref{sec_JobSN} and \ref{sec_RepSN}).

The sole use of SRP does not allow for comparing entities with a distance $\leq w$ that spread over different reducers. For example, standard SN determines the correspondence $(h,c)$ (see Figure~\ref{fig_sorted_neighborhood}) that can not be generated since $h$ and $c$ are assigned to different reducers. For $r$ reducers and a window size $w$, SRP misses $(r-1)\cdot w \cdot (w-1) /2$ boundary correspondences. We therefore present two approaches, JobSN and RepSN, that build on SRP but are also able to deal with boundary entities.

\subsection{JobSN: Sorted Neighborhood with additional MapReduce job}
\label{sec_JobSN}

\begin{figure}[t]
  \begin{center}
    \includegraphics[scale=0.435]{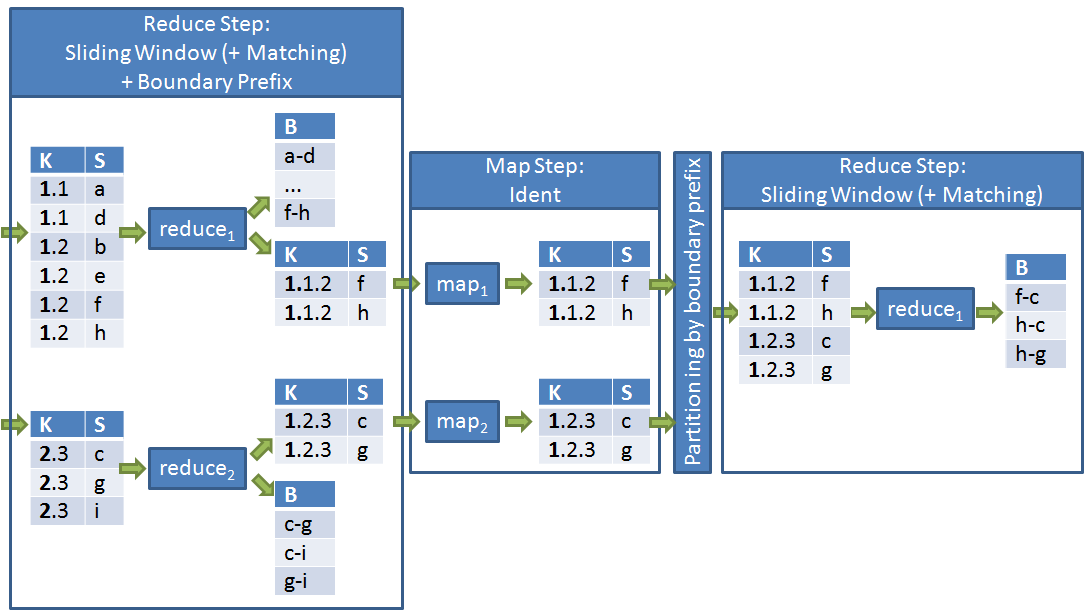}
    \caption{\label{fig_sn_additional_job}Example execution of SN with additional MapReduce job (JobSN, $w=3$). The far left box is the reduce step of the first job. Its output is the input to the second MapReduce job.}
  \end{center}
\end{figure}

The JobSN approach utilizes SRP and employs a second MapReduce job afterwards that completes the SN result by generating the boundary correspondences. JobSN makes thereby use of the fact that MapReduce provides sorted partitions to the reducer. A reducer can therefore easily identify the first and the last $w-1$ entities during the sequential execution. Those entities have counterparts in neighboring partitions, i.e., the last $w-1$ entities of a reducer relate to the first $w-1$ entities of the succeeding reducer. In general, all reducers output the first and last $w-1$ entities with the exception of the first and the last reducer. The first (last) reducer only returns the last (first) $w-1$ entities.

The pseudo-code for JobSN is shown in the appendix in Algorithm~\ref{alg_JobSN}. Figure~\ref{fig_sn_additional_job} illustrates a JobSN execution example. It uses the same data of Figure~\ref{fig_sn_mr}. The map step of the first job is identical with SRP of Figure~\ref{fig_sn_mr} and omitted in Figure~\ref{fig_sn_additional_job} to save space. The reduce step is extended by an additional output. Besides the list of blocking correspondences $B$, the reducer also emits the first and last $w-1$ entities.

JobSN realizes the assignment of related boundary elements with an additional boundary prefix that specifies the boundary number. Since the last $w-1$ entities of reducer $i<r$ refer to the $i^{th}$ boundary, the keys of the last $w-1$ entities are prefixed with $i$. On the other hand, the first $w-1$ entities of the succeeding reducer $i+1$ also relate to the $i^{th}$ boundary. Therefore the keys of the first $w-1$ entities of reducer $i>1$ are prefixed with $i-1$. The first reducer in the example of Figure~\ref{fig_sn_additional_job} prefixes the last entities ($f$ and $h$) with 1 and the second reducer prefixes the first entities ($c$ and $g$) with 1, too. Thereby the key reflects data lineage: The actual blocking key of entity $c$ is 3 (see, e.g., Figure~\ref{fig_sorted_neighborhood}), it was assigned to reducer number 2 during the SRP (Figure~\ref{fig_sn_mr}), and it is associated with boundary number 1 (Figure~\ref{fig_sn_additional_job}).

The second MapReduce job of JobSN is straightforward. The map functions leaves the input data unchanged. The map output is then redistributed to the reducers based on the boundary prefix. The reduce function then applies the sliding window but filters correspondences that have already been determined in the first MapReduce job. For example, $(f,h)$ does not appear in the output of the second job since this pair is already determined by SRP. As mentioned above, this knowledge is encoded in the lineage information of the key because those entities share the same partition number.

The JobSN approach generates the complete SN result at the expense of an additional MapReduce job. We expect the overhead for an additional job to be acceptable and we will evaluate JobSN's performance in Section~\ref{sec_experiments}.

\subsection{RepSN: Sorted Neighborhood with entity replication}
\label{sec_RepSN}

\begin{figure}[t]
  \begin{center}
    \includegraphics[scale=0.45]{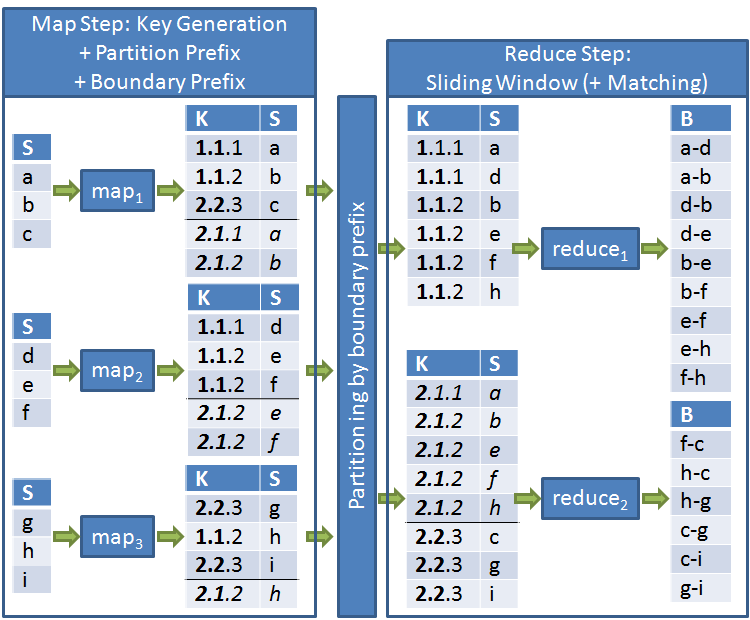}
    \caption{\label{fig_sn_replication}Example execution of sorted neighborhood with entity replication (RepSN, $w=3$). Entities are replicated within in the map function (below solid line). Replicated entities are written in \textit{italic type}.}
  \end{center}
\end{figure}

The RepSN approach aims to realize SN within a single MapReduce job. It extends SRP by the idea that each reducer $i>1$ needs to have the last $w-1$ entities of the preceding reducer $i-1$ in front of its input. This would ensure that the boundary correspondences appear in the reducer's output. Unfortunately, the MapReduce paradigm is not designed for mutual data access between different reducers. MapReduce only provides options for controlled data replication within the map function.

The RepSN approach therefore extends the original SRP map function so that map replicates an entity that should be send to both the respective reducer and its successor. For all but the last reduce partition $r$, the map function thus identifies the $w-1$ entities with the highest blocking key $k$. It first outputs all entities and adds the identified boundary entities afterwards. Similar to SRP, an entity key is determined by the blocking key $k$ plus a partition prefix $p(k)$. To distinguish between original entities and replicated boundary entities, RepSN adds an additional boundary prefix. For all original entities this boundary prefix is the same as the partition number, i.e., the composite key is $p(k).p(k).k$. The boundary prefix for replicated entities is the partition number of the succeeding reducer, i.e., the composite key is $(p(k)+1).p(k).k$.


RepSN is described in the appendix in Algorithm~\ref{alg_RepSN}. Figure~\ref{fig_sn_replication} illustrates an example execution of RepSN. The example employs $r=2$ reducers and window size $w=3$. Therefore all mappers identify the $w-1=2$ entities with the highest key of partition 1. The output of each map function is divided into two parts. The upper part (above the solid line) is equivalent to the regular map output of SRP. The only (technical) difference is that the partition prefix is duplicated. The lower part (below the solid line) of the map output contains the replicated entities. Consider the second map function: All three entities ($d$, $e$, and $f$) are assigned to the partition 1 and $e$ and $f$ are replicated because they have the highest keys. The keys of the replicated data start with the succeeding partition 2. This ensures that $e$ and $f$ are send to both reducer 1 and reducer 2.

The map output is then redistributed to the reduce functions based on the boundary prefix. Furthermore, MapReduce provides a sorted list as input to the reduce functions. Due to the structure of the composite key, the replicated entities appear at the beginning of each reducer input. Replicated entities share the same boundary prefix but have a smaller partition prefix. The reduce function then applies the sliding window approach but only returns correspondences involving at least one entity of the actual partition.

In the example of Figure~\ref{fig_sn_replication}, input and output of the first reducer are equivalent to SRP (see Figure~\ref{fig_sn_mr}). The second reducer receives a larger input partition. It ignores all replicated entities but the $w-1=2$ highest ($f$ and $h$). The output is the union of the corresponding SRP output and the corresponding boundary reduce output of JobSN.

RepSN allows for an entire sorted neighborhood computation within a single MapReduce job at the expense of some data replication. Since the MapReduce model does not provide any global data access during the computation, it is not possible to identify only the necessary entities for processing the boundary elements. Rather each map function has to identify and replicate possibly relevant entities based on its local data. Each mapper has to replicate $w-1$ entities for all but the last partition. The maximum number of replicated entities is therefore $m \cdot (r-1) \cdot (w-1)$. This number is independent from the size $n$ of input entities and may therefore be comparatively small for large datasets. We will evaluate the overhead of data replication and data transfer in Section~\ref{sec_experiments}. In particular we will compare it against the JobSN overhead for scheduling and executing an additional MapReduce job.

\section{Experiments}
\label{sec_experiments}
We conducted a set of experiments to evaluate the effectiveness of the proposed approaches. After a description of the experimental setup we study the scalability of our Sorted Neighborhood approaches. Afterwards we will discuss the effects of data skew and show its influence on execution time.

\subsection{Experimental setup}
We run our experiments on up to four nodes with two cores. Each node has an Intel(R) Core(TM)2 Duo E6750 2x2.66GHz CPU, 4GB memory and runs a 64-bit Debian GNU/Linux OS with a Java 1.6 64-bit server JVM. On each node we run Hadoop 0.20.2. Following \cite{setsim_mr} we made the following changes to the Hadoop default configuration: We set the block size of the DFS to 128MB, allocated 1GB to each Hadoop daemon and and 1GB virtual memory to each map and reduce task. Each node was configured to run at most two map and reduce tasks in parallel. Speculative execution was turned off. Both master daemons for managing the MapReduce jobs and the DFS run on a dedicated server. We used Hadoop's SequenceFileOutputFormat with native bzip2 block compression to serialize the output of mappers and reducers that was further processed.  Sequence files can hold binary $(key,value)$ pairs what conceptually allowed us to deal with $(String,String[])$ instead of $(String,String)$ pairs. Hence, we could directly access the \textit{i}$^{th}$ attribute value of an entity during matching in comparison to split a string at runtime.

The input dataset\footnote{http://asterix.ics.uci.edu/data/csx.raw.txt.gz} for our experiments contains about 1.4 Mio. publication records. To compare two publications we executed two matchers (edit distance on title, TriGram on abstract) and calculated a weighted average of the two results. Pairs of entities with an average similarity score of at least 0.75 were regarded as matches. We applied an internal optimization by skipping the execution of the second matcher if the similarity after the execution of the first matcher was too low for reaching the combined similarity threshold. To group similar entities into blocks we used the lowercased first two letters of the title as blocking key.

\subsection{Sorted Neighborhood}
\label{sec_exp_sn}
We evaluate the absolute runtime and the relative speedup using two window sizes of 10 and 1000. The additional MapReduce job of JobSN was executed with one reducer ($r=1$). To ensure comparability for different numbers of mappers and reducers we used the same manually defined function in each experiment partitioning the entities into 10 blocks of slightly varying size (e.g., many publication titles start with ``a ''). The resulting 10 reduce tasks are executed by at most 8 reducers (see Section~\ref{sec_skew} for further discussion).

\begin{figure}
  \subfigure[window size $w=10$]{\label{fig_sn_eval_10}\includegraphics[width=0.49\textwidth]{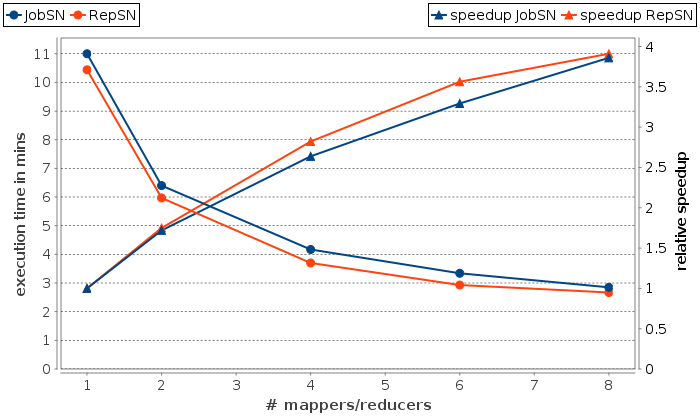}}\hfill
  \subfigure[window size $w=1000$]{\label{fig_sn_eval_1000}\includegraphics[width=0.49\textwidth]{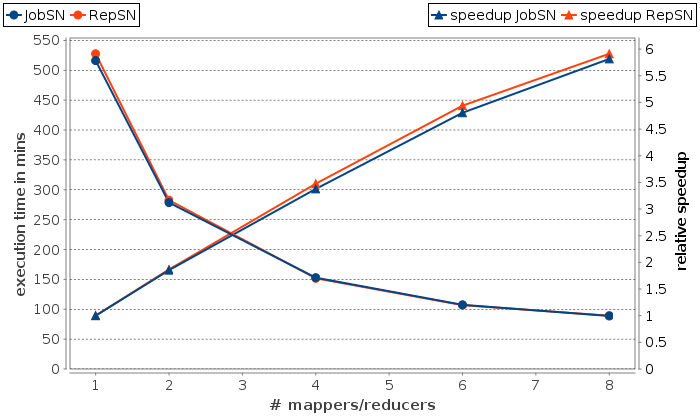}}
  \caption{\label{fig_sn_eval}Comparison of the two Sorted Neighborhood implementations}
\end{figure}

Figure~\ref{fig_sn_eval} shows execution times and speedup results for up to 8 mappers and 8 reducers for the two proposed implementations.  The configurations with $m=r=1$ refers to sequential execution on a single node, the one with $m=r=2$ refers to the execution on a single node utilizing both cores and so on. For the small window size $w=10$, RepSN slightly outperforms JobSN due to the scheduling overhead of JobSN for the additional MapReduce job. The execution time of both RepSN and LocSN could be reduced from approximately 10.5 to about 2.5-3 minutes resulting in a relative speedup of up to 4 for 8 cores. For the larger window size $w=1000$, the execution times scale linearly, for instance the execution time for RepSN could be reduced from approximately 9 to merely 1.5 hours. We observe a linear speedup for the entire range of up to 4 nodes and 8 cores. The runtime of the different implementations differ only slightly. Differences can only be observed for a small amount of parallelism, i.e., RepSN was 10 minutes slower for $w=1000$ in the sequential case. Beginning with $m=r=4$ RepSN completed faster than JobSN.  The reasons for the suboptimal speedup values (about 6 for 8 cores) are caused by design and implementation choices of MapReduce/Hadoop to achieve fault tolerance, e.g., materialization of (intermediate) results between map and reduce.


\subsection{Data skew}
\label{sec_skew}
We finally study the effects of data skew and use RepSN for this experiment. Practical data skew handling has been studied in the context of parallel DBMS \cite{practial_skew_handling} but has not yet been incorporated in our implementation. The application of the partitioning function $p$ as described in Section~\ref{sec_LocSN} is susceptible to data skew and resulting load imbalances. This is because partitions may be of largely varying size so that the total execution time is dominated by a single or few reducers.

\begin{figure}[tb]
  \begin{minipage}[t]{0.35\textwidth}
    \centering
    \raisebox{\depth}{
      \begin{tabular}{|l||c|c|}
	\hline \textbf{p} & \textbf{g} \\ \hline \hline
	Manual & 0.13 \\ \hline
	Even10 & 0.30 \\ \hline
	Even8 &  0.32 \\ \hline
	Even8\_40 & 0.42 \\ \hline
	Even8\_55 & 0.54 \\ \hline
	Even8\_70 & 0.63 \\ \hline
	Even8\_85 & 0.76 \\ \hline
      \end{tabular}
    }
    \captionof{table}{Partitioning functions and resulting data skew}
    \label{tab_skew}
  \end{minipage}
  \hfill
  \begin{minipage}[t]{0.49\textwidth}
    \centering
    \includegraphics[width=1\textwidth]{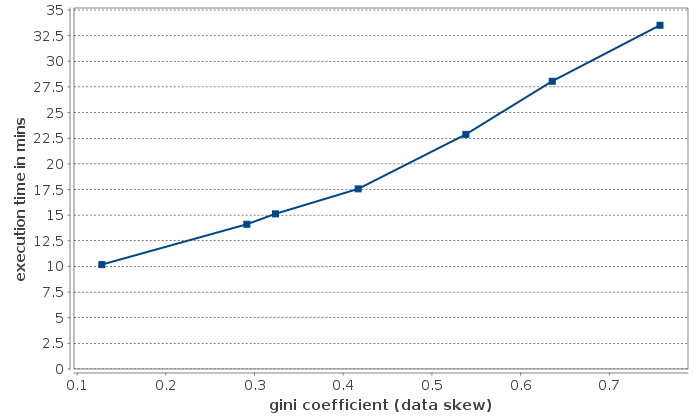}
    \caption{Execution time of RepSN for various degrees of data skew ($w=100$)}
    \label{fig_skew}
  \end{minipage}
  \caption{Influence of data skew ($m=r=8$)}
\end{figure}

We ran our experiments on all 4 nodes (8 mappers and 8 reducers) with a window size $w=100$. To quantify the inequality of the key distribution in the dataset we utilize the Gini coefficient $g=\frac{2\cdot \sum^n_{i=1} i\cdot y_i}{n\cdot \sum_{i=1}^n y_i} - \frac{n+1}{n}$, whereas $y_i$ is the number of entities in partition $i$ and $y_i \leq y_{i+1}$. A value of 0 expresses total equality and a value of 1 maximal inequality. We evaluated the partition strategies shown in Table~\ref{tab_skew} exhibiting different degrees of data skew as indicated by their gini coefficient. In addition to the manually defined partitioning function used in Section~\ref{sec_exp_sn} with almost equally-sized partitions we evenly partitioned the key space into 10 and 8 intervals (Even10, Even8). Finally we used Even8 but modified the blocking keys so that 40\%, 55\%, 70\% and 85\%, respectively, of all entities fall in the last partition. The runtime results for the different partitioning strategies are illustrated in Figure~\ref{fig_skew}. The manual partitioning strategy that was tuned for equally-sized partitions performed best, while the most skewed configuration suffered from a more than threefold execution time.  Even10 completed slightly (one minute) faster than Even8 because of its smaller partitions allowing the 8 reducers processing  several small partitions while a large partition is matched (improved load balancing potential). For Even8\_40 - Even8\_85 we observe significant increases of the execution time with a rising degree of data skew. Clearly the influence of data skew will increase for larger window sizes since more entities within a partition have to be compared bye one reducer.

The observed problems are MapReduce-inherent because the programming model demands that all values with the same key are processed by the same reducer. A majority of values for one or a small subset of the dataset's keys does not allow effective parallel data processing. There is no load balancing mechanism in MapReduce except the redundant execution of outstanding map or reduce task at the end of a job (speculative execution). However, this helps only to deal with partially working or misconfigured stragglers. Therefore it becomes necessary to investigate in load balancing mechanisms for the MapReduce paradigm.

\section{Related work}
\label{sec_related_work}

Entity resolution is a very active research topic and many approaches have been proposed and evaluated as described in recent surveys \cite{er_survey, er_frameworks}. Surprisingly, there are only a few approaches that consider parallel entity resolution. First ideas for parallel matching were described in the Febrl system \cite{pfebrl_and_ngram}. The authors show how the match computation can be parallelized among available cores on a single node. Parallel evaluation of the Cartesian product of two sources considering the three input cases (clean-clean, clean-dirty, dirty-dirty) is described in \cite{parallel_linkage}.

\cite{data_partitioning} proposes a generic model for parallel processing of complex match strategies that may contain several matchers. The parallel processing is based on general partitioning strategies that take memory and load balancing requirements into account. Compared to this work \cite{data_partitioning} allows the execution of a match workflow on the Cartesian product of input entities. This is done by partitioning the set of input entities and generating match tasks for each pair of partitions. A match task is then assigned to any idle node in a distributed match infrastructure with a central master node. The advantage of this approach is the high flexibility for scheduling match tasks and thus for dynamic load balancing. The disadvantage is that only the matching itself is executed in parallel. Blocking is done upfront on the master node. Furthermore in this work we rely on an widely used  parallel processing framework that hides the details of parallelism and therefore is less error-prone.

We are only aware of one previous approach for parallel entity resolution on a cloud infrastructure \cite{setsim_mr}. The authors do not investigate Sorted Neighborhood blocking but show how a single token-based string similarity function can be realized with MapReduce. The approach is based on a complex workflow consisting of several MapReduce jobs. This approach suffers from similar load balancing problems as observed in Section~\ref{sec_skew} because all entities that share a frequent token are compared by one reducer. In contrast to our Sorted Neighborhood approach large partitions for frequent tokens that do not fit into memory must be handled separately. This is because all entities that contain a specific token have to be compared with each other instead of comparing only entities with a maximum distance of less than $w$. Compared to \cite{setsim_mr}, we are not limited to a specific similarity function but can apply a complex match strategy for each pair of entities within a window. Furthermore as explained in Section 3 Sorted Neighborhood can be substituted with other blocking techniques, e.g., Standard Blocking or N-gram indexing.

\section{Conclusions and outlook}
We have shown how entity resolution workflows with a blocking strategy and a match strategy can be realized with MapReduce. We focused on parallelizing Sorted neighborhood blocking and proposed two MapReduce-based implementations. The evaluation of our approaches demonstrated their efficiency and scalability in comparison to sequential entity resolution. We also pointed out the need for incorporating load balancing and skew handling mechanisms with MapReduce.

There are further limitations of MapReduce and the utilized implementation Hadoop such as insufficient support for pipelining intermediate data between map and reduce jobs.  There are other parallel data processing frameworks like \cite{nephele} that support different types of communication channels (file, TCP, in-memory) and provide a better support for different input sets. Furthermore there are concepts like \cite{map_reduce_merge} that propose to adapt and extend MapReduce to simplify set operations (Cartesian product) on heterogeneous datasets.

In future work we plan to investigate load balancing and data partitioning mechanisms for MapReduce.

\phantomsection
\addcontentsline{toc}{section}{References}
\bibliography{references}

\appendix
\section{Algorithms}

Algorithm~\ref{alg_JobSN} and Algorithm~\ref{alg_RepSN} show the pseudo-code for the two proposed Sorted Neighborhood implementation JobSN and RepSN introduced in sections~\ref{sec_JobSN} and ~\ref{sec_RepSN}. For simplicity, we use a function \verb!StandardSN! that implements the standard Sorted Neighborhood approach, i.e., that moves the window of size $w$ over a sorted list of entities and outputs matching entity pairs (Algorithm~\ref{alg_JobSN} line~\ref{line_SN_JobSN}, \ref{line_SN_JobSN2} and Algorithm~\ref{alg_RepSN} line~\ref{line_SN_RepSN}).

Throughout the two algorithms $r$ denotes the number configured number of reducers for the MapReduce job. The partitioning function $p: k \rightarrow i$ with $1 \leq i \leq r$ determines the reducer $r_i$ to which an entity with the blocking key value $k$ is repartitioned. A key of the form $x.y$ denotes a composed key of $x$ and $y$. Composed keys are compared component-wise. The comments indicate which parts of the composite keys are used for map-side repartitioning and reduce-side grouping of entities.

For simplicity, the pseudo-code of Algorithm~\ref{alg_JobSN} does not filter correspondences that have been already determined in the first phase. In Algorithm~\ref{alg_RepSN} we use two extra functions in addition to map and reduce. The function $map\_configure$ is executed before a mapper executes a map task and $map\_close$ before termination of a map task, respectively.

 \newpage
\begin{algorithm}[H]
\small
\SetKwFunction{map}{map}
\SetKwFunction{reduce}{reduce}
\SetKwFunction{SN}{StandardSN}
\SetKwFunction{output}{output}
\SetKwBlock{Begin}{}{}
\SetKwData{K}{k}
\SetKwData{RI}{r$_i$}
\SetKwData{bound}{bound}
\SetKwData{first}{first}
\SetKwData{last}{last}

\tcp{ --- Phase 1 ---}
\map{key$_{in}$=unused, value$_{in}$=entity}
\Begin{
  \K $\leftarrow$  generate blocking key for \ArgSty{entity}\;
  \RI $\leftarrow$  p(\K) \tcp*[l]{reducer to which entity is assigned by p}
  \tcp{Use composite key to partition by r$_i$}
  \output{key$_{tmp}$=\RI.\K, value$_{tmp}$=entity}
}

\BlankLine 
\tcp{group by r$_i$, order by composed key}
\reduce{key$_{tmp}$=\RI.\K, list(value$_{tmp}$)=list(entity)}
\Begin{
  \SN{list(entity), w}\; \label{line_SN_JobSN}
  \first $\leftarrow$ first $w-1$ entities of \ArgSty{list(entity)}\;
  \last $\leftarrow$ last $w-1$ entities of \ArgSty{list(entity)}\;
  \If{r$_i>$ 1}{
    \bound $\leftarrow$\RI-1\;
    \ForEach{entity $\in$ \first}{
      \output{key$_{out}$=\bound.\RI.\K, value$_{out}$=entity}
    }
  }

  \If{r$_i<$ r}{
    \bound $\leftarrow$ \RI\;
    \ForEach{entity $\in$ \last}{
      \output{key$_{out}$=\bound.\RI.\K, value$_{out}$=entity}
    }
  }
}

\BlankLine
\BlankLine
\tcp{ --- Phase 2 ---}
\map{key$_{in}$=\bound.\RI.\K, value$_{in}$=entity}
\Begin{
  \tcp{Use composite key to partition by bound}
  \output{key$_{tmp}$=\bound.\RI.\K, value$_{tmp}$=entity}
}

\BlankLine 
\tcp{group by bound, order by composed key}
\reduce{key$_{tmp}$=\bound.\RI.\K, list(value$_{tmp}$)=list(entity)}
\Begin{
   \SN{list(entity), w}\;\label{line_SN_JobSN2}
}
\caption{\label{alg_JobSN}JobSN}
\end{algorithm}

\newpage

\begin{algorithm}[H]
\small
\SetKwFunction{mapConfigure}{map\_configure}
\SetKwFunction{map}{map}
\SetKwFunction{mapClose}{map\_close}
\SetKwFunction{reduce}{reduce}
\SetKwFunction{SN}{StandardSN}
\SetKwFunction{output}{output}
\SetKwBlock{Begin}{}{}
\SetKwData{K}{k}
\SetKwData{RI}{r$_i$}
\SetKwData{Rep}{rep}
\SetKwData{bound}{bound}
\SetKwData{Min}{min}
\SetKwData{MinKey}{k$_{min}$}

\mapConfigure
\Begin{
  \tcp{list of w-1 entities with the highest}
  \tcp{blocking key for partition i<r}
  \ForEach{i $\in \lbrace 1, \ldots, r-1\rbrace$}{
    \Rep$_{i}$ $\leftarrow$ [];
  }
}

\BlankLine
\map{key$_{in}$=unused, value$_{in}$=entity}
\Begin{
  \K $\leftarrow$  generate blocking key for \ArgSty{entity}\;
  \RI $\leftarrow$  p(\K) \tcp*[l]{reducer to which entity is assigned by p}
  \bound $\leftarrow$ \RI\;
  \If{r$_i<$ r}{
    \If{sizeOf(\Rep$_{r_i}$)$<$w-1}{
      append(\Rep$_{r_i}$, \ArgSty{entity});
    }
    \Else{
      \Min $\leftarrow$ determine entity from \Rep$_{r_i}$ with smallest blocking key\;
      \MinKey $\leftarrow$ blocking key of \Min\;
      \If{\K$>$\MinKey}{
	replace(\Rep$_{r_i}$, \Min, \ArgSty{entity});
      }
    }
  }
  \BlankLine
  \tcp{Use composite key to partition by bound}
  \output{key$_{tmp}$=\bound.\RI.\K, value$_{tmp}$=entity}
}

\BlankLine
\mapClose
\Begin{
  \ForEach{i $\in \lbrace 1, \ldots, r-1\rbrace$}{
    \RI $\leftarrow$ \ArgSty{i}\;
    \bound $\leftarrow$ \RI+1\;
    \ForEach{entity $\in$ \Rep$_i$}{
      \tcp{prefix key with r$_i$+1 to assign replicated}
      \tcp{entities to succeeding reducer}
      \output{key$_{tmp}$=\bound.\RI.\K, value$_{tmp}$=entity}
    }
  }
}

\BlankLine
\tcp{group by bound, order by composed key}
\reduce{key$_{tmp}$=\bound.\RI.\K, list(value$_{tmp}$)=list(entity)}
\Begin{
  remove all entities with \bound $\neq$ \RI from the head of \ArgSty{list(entity)} except the last $w-1$\;
  \SN{list(entity), w}\;\label{line_SN_RepSN}
}
\caption{\label{alg_RepSN}RepSN}
\end{algorithm}

\end{document}